\documentclass[preprint,showpacs,preprintnumbers,amsmath,amssymb,floats]{revtex4}
\usepackage[dvips]{graphicx}
\usepackage[english]{babel}
\usepackage{amsmath}
\usepackage{amssymb}
\usepackage{color}

\usepackage{subfig}
\usepackage{leftidx}
\usepackage{bm}

\newcommand{\be}{\begin{eqnarray}}
\newcommand{\ee}{\end{eqnarray}}

\begin{document}
\title{Zeeman Modulated Spin Echo in Orthorhombic Symmetry}\author{Changtao Hou, D.E. MacLaughlin and C.M. Varma}
\affiliation{Department of Physics, University of California, Riverside, CA}
\date{\today}
\begin{abstract}
The experimental study of the modulation of the envelope of spin-echo signals due to internal and external fields is an important spectroscopic tool to detect very small internal magnetic fields. We derive the free induction decay and the frequency spectrum and amplitude of spin-echo signals for arbitrary orientation of fields with respect to crystalline axis for nuclei in a crystal of orthorhombic symmetry. Results reproduce the results that no modulation should be observed in tetragonal crystals for fields either along the c-axis or any direction in the basal plane and give details of the signal as a function of the orthorhombicity parameter. They are used to discuss recent experimental results and provide guidelines for future experiments.
\end{abstract}
\maketitle

\section{Introduction}
Spin-echo experiments \cite{Abragam, Slichter} are a principal method to study magnetic fields and magnetic relaxation processes in liquids and solids. In the conventional spin-echo experiment, rf pulses orthogonal to a uniform external magnetic field ${\bf H_0}$ are applied and the time dependence of the free induction signal following a single pulse and the amplitude of the spin-echo signal as a function of the separation of two applied rf pulses with frequency near the nuclear resonance frequency due to the Zeeman splitting of levels by ${\bf H_0}$ are studied. In crystals in which the nuclear levels of some ions have quadrupolar splittings, the echo technique can be used at zero uniform external field. Small uniform fields applied often split the quadrupole levels in such systems and yield additional modulations of envelope of the echo signals  at frequencies which  depend on the splittings.  This technique may also be used as a very sensitive method to study the direction and magnitude of small internal magnetic fields in a solid. This technique was also invented long ago \cite{Hahn}. The theory of the Zeeman modulation of Quadrupolar echoes was fully described for crystals with tetragonal symmetry by Das and Saha \cite{Das1} and \cite{bloom} and has been reviewed by Das and Hahn \cite{Das2}.

The Zeeman modulated spin-echo  technique has been used to study \cite{Lombardi1} internal magnetic fields at $^{137}$Ba nuclei in powder samples of insulating AFM YBa$_2$Cu$_3$O$_{6.05}$, which is a tetragonal crystal. In the former case oscillatory modulations of Spin-echo envelope are observed but with a frequency about half of what the authors expect on the basis of theoretical expressions that they state without derivation, but which surprisingly are not in agreement with the standard published results \cite{Das1, Das2, bloom}. The standard published theoretical results for the tetragonal crystals unambiguously state that {\it no modulation} is to be expected for Zeeman fields along the tetragonal axis or any direction in the basal plane of a tetragonal crystal. This result does not change for {\it internal} fields on Ba nuclei inside the crystallites in a poly-crystalline or powder sample. (Small effects might occur for nuclei sitting in ill defined symmetries at interfaces.) The obtained experimental results cannot therefore be explained by the theory which is effectively exact.

The same authors \cite{Mali} have looked for internal fields  expected  due to the magnetic order proposed \cite{cmv} and observed by polarized neutron scattering \cite{Bourges, Greven} and dichroic Arpes \cite{kaminski} in the under-doped cuprates. Results consistent with zero internal fields are observed at the Ba nuclei in the under-doped cuprate YBa$_2$Cu$_4$O$_8$. The authors again use their own theoretical expression for tetragonal crystals to compare with experiments.  As mentioned already, according to the standard results, they should not observe any oscillation in a tetragonal crystal. So, at this point, one could say that the authors see precisely what is to be expected on the basis of the well described theory. But YBa$_2$Cu$_4$O$_8$ is orthorhombic.  The authors state that ortho-rhombicity should not matter.  We find that this also is not correct. Although some results for the orthorhombic symmetry with the correct conclusion have been stated \cite{ZS}, no clear theoretical derivation, at the same level as available for tetragonal symmetry, is available. Our aim in the present work is to provide such a derivation.

With the derived results, one can put lower limits on the orthorhombic splitting of the quadrupolar lines for the spin-echo modulation to be observed at a given internal field. Estimates of such splittings are available from experiments. Our conclusion is that given those values the modulation should have indeed been observed if the internal fields were static. Similarly no signals due to the static magnetic fields are observed in under-doped cuprates in muon \cite{Mcdoo} experiments or in more direct NMR experiments \cite{Halperin}. On the other hand, polarized neutron scattering diffraction experiments have observed the predicted order in four different families of cuprates \cite{Bourges}. A large birefringence effect \cite{Armitage} consistent with the predicted order has also been observed  \cite{cmv-GB} in under-doped cuprates. A reconciliation of these experiments is possible. In a new development \cite{cmv2}, it is  argued that the observed order by neutrons can not be truly static given the disorder in the crystals which induces domain formations with lengths of order $10^2$ Angstroms. Neutron scattering experiments integrate over frequencies of $O(10^11 Hz)$, while the NQR (and muon) experiments look for signals at frequencies of $O(10^7 Hz)$. It is suggested \cite{cmv2} that the finite frequency fluctuations in domains of finite size notionally narrow the signal so that no effects in NQR or muons are observed.  

\section{Quadrupolar Hamiltonian with  Zeeman-perturbation}
The Hamiltonian for a quadrupole in an orthorhombic crystalline field in the presence of a perturbing Zeeman field is 
\be
\label{fullH}
H_0=H_Q+H_Z
\ee
where
\be
H_Q=\frac{e^2qQ}{4I(2I-1)}[3I_z^2-I^2+\eta(I_x^2-I_y^2)].
\ee
 $\eta$ is the ortho-rhombicity parameter, and the Zeeman Hamiltonian,
\be
H_Z=-\hbar\gamma\bm{H_0}\cdot\bm{I}.
\ee
 For $\bm{H_0} =0$, the two pairs of doubly degenerate energy levels for $I=3/2$ are given by
\be
\label{ev1}
E_{1,2}=-\frac{e^2qQ}{4}\rho,~~
E_{3,4}=\frac{e^2qQ}{4}\rho
\ee
where $\rho=\sqrt{1+\eta^2/3}$. \\
The orthonormal wavefunctions associated with these levels are given by 
\be
\label{basis}
|\phi_1\rangle&=&\cos\chi|1/2\rangle-\sin\chi|-3/2\rangle,\nonumber\\
|\phi_2\rangle&=&\cos\chi|-1/2\rangle-\sin\chi|3/2\rangle,\nonumber\\
|\phi_3\rangle&=&\cos\chi|3/2\rangle+\sin\chi|-1/2\rangle,\nonumber\\
|\phi_4\rangle&=&\cos\chi|-3/2\rangle+\sin\chi|1/2\rangle
\ee
where $\sin\chi=\sqrt{\frac{\rho-1}{2\rho}}$ and $|\pm1/2\rangle$ and $|\pm3/2\rangle$ are the eigenfunctions of $I_z$. The representation of the quadrupole operators $I_x, I_y, I_z$, in this basis, necessary for the calculation of the spin-echoes are given in the Appendix.

\begin{figure}
\includegraphics[width=0.5\columnwidth]{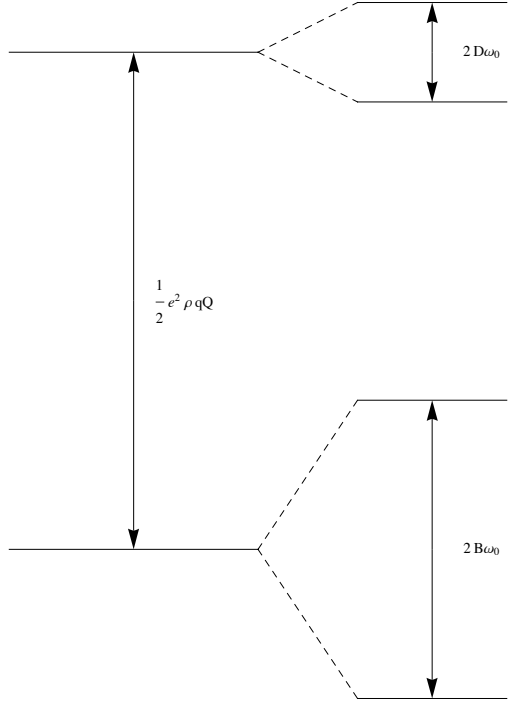}
\caption{The energy levels due to quadrupole energy and Zeeman energy. The parameters B and D and the wave-function for each of the levels is given in the text.}
\label{Fig_energysplit}
\end{figure}

We are interested in the the case $H_Q\gg H_Z$. Then the wavefunctions and the energy levels can be obtained by treating the Zeeman term as a perturbation. Details of this derivation are given in the Appendix and we show only the final results here. We are interested in the general case that $\bm{H_0}$ is oriented at an angle $\theta_0$ with respect to the crystalline c-axis and makes an angle $\phi_0$ with respect to the crystalline a-axis. The degeneracy of the energy levels is split and the eigenvalues may be written compactly as 
\be
\label{ev2}
E_{2,1}&=&-E_Q\pm\hbar D\omega_0,\nonumber\\
E_{4,3}&=&E_Q\pm\hbar B\omega_0.
\ee
Fig. (\ref{Fig_energysplit}) shows the energy level scheme. The quantum numbers of the 4 levels are $I_z = \pm 1/2, \pm 3/2$ only when the Zeeman field is along the symmetry axis $\hat{z}$. Otherwise they are mixed as discussed below. The mixed wave-functions are crucial in determining the transitions in an rf field and therefore the pattern of oscillations of the quadrupole echoes.

The important parameters $B$ and $D$ are given by
\be
\label{BD}
D&=&\frac{[(2-\rho)^2\cos^2\theta_0+(\rho+1-\eta)^2\sin^2\theta_0\cos^2\varphi_0+(\rho+1+\eta)^2\sin^2\theta_0\sin^2\varphi_0]^{1/2}}{2\rho},\nonumber\\
B&=&\frac{[(2+\rho)^2\cos^2\theta_0+(\rho-1+\eta)^2\sin^2\theta_0\cos^2\varphi_0+(\rho-1-\eta)^2\sin^2\theta_0\sin^2\varphi_0]^{1/2}}{2\rho}
\ee
The first order orthonormal eigenfunctions are given by,
\be
\label{basis2}
|\xi_1\rangle&=&-d^\ast|\phi_2\rangle+c^\ast|\phi_1\rangle,\nonumber\\
|\xi_2\rangle&=&c|\phi_2\rangle+d|\phi_1\rangle,\nonumber\\
|\xi_3\rangle&=&-b^\ast|\phi_4\rangle+a^\ast|\phi_3\rangle,\nonumber\\
|\xi_4\rangle&=&a|\phi_4\rangle+b|\phi_3\rangle.
\ee
The coefficients in (\ref{basis2}) are given in the Appendix.

\section{Time-development of the wave-function}
Let $H_1$ specify the Hamiltonian for the rf field coupling to the quadrupole. 
\be
\label{rf}
H_1&=&-\hbar\gamma\bm{H_r}\cdot\bm{I}\cos(\omega t), \\
\bm{H_r}&=&H_r(\sin\theta_1\cos\varphi_1,\sin\theta_1\sin\varphi_1,\cos\theta_1).
\ee
In the experiments, this field is applied in two very short pulses each with width $t_w$ and time $\tau$ between them. To calculate the spin-echo, all one needs is to calculate the time-development of the wave-function (in the basis of Eq. ({\ref{basis}) with and without the rf-pulses applied. Let $R(t)$ and $D(t)$ be the time-evolution operator in the presence and absence of the rf pulses. $R(t)$ and $D(t)$ are then given by
\be
\label{time-dev}
i\hbar\frac{dR}{dt}=(H_0+H_1)R, ~~
i\hbar\frac{dD}{dt}=H_0D
\ee
The density matrix $\rho(t)$ for the spin system after the passage of the pulses is related to the initial density matrix $\rho(0)$ before the pulses are applied by the relation
\begin{eqnarray}
\rho(t)=S(t)\rho(0)S^{-1}(t), ~
S=D(t-\tau-t_w)R(t_w)D(\tau-t_w)R(t_w).
\ee
 For the initial condition, i.e. at time $t=0$, the density matrix $\rho(0)$ is given by 
$exp(-H_Q/k_BT)$, because the Zeeman field part $H_Z$ is included in the time-development. Also, the energy differences due to$H_Q$ are much smaller than $k_BT$, so only the leading term in $H_Q/k_BT$ need be kept. Therefore,
\be
\rho(0)={\bf 1} -\frac{-e^2qQ}{4k_BT}\left(\begin{array}{cccc}
\rho&0&0&0\\
0&\rho&0&0\\
0&0&-\rho&0\\
0&0&0&-\rho
\end{array}\right).
\ee

\noindent
In most experiments, where the rf coil producing the rf field is also the detector, we need the expectation value of $I_x(t)$ after the passage of pulses. This is given by
\begin{eqnarray}
\label{I}
\langle I_i\rangle=Tr\lbrace\rho(t)I_i\rbrace=Tr\{S\rho(0)S^{-1}I_i\}.
\end{eqnarray}

\section{General Expression for the Nuclear Quadrupole Spin Echo Envelope Modulations}
In the absence of the r.f. pulses, the Hamiltonian is independent of time and the evolution is governed by the time evolution operator(set $\hbar=1$) 
\be
D(t-t_0)=\exp[-iH_0(t-t_0)].
\ee
With the r.f. pulse, we need to work in the interaction representation defined by the transformation $U_0=\exp[-iH_Qt]$.  $H_1$ in $|\phi_i\rangle$ basis is
\be
H_1=\omega_1\left(\begin{array}{cccc}
0&0&\lambda_x-i\lambda_y&-\lambda_z\\
0&0&\lambda_z&\lambda_x+i\lambda_y\\
\lambda_x+i\lambda_y&\lambda_z&0&0\\
-\lambda_z&\lambda_x-i\lambda_y&0&0
\end{array}\right),
\ee
where
\be
\lambda_x &=&\frac{\eta+3}{2\sqrt{3}\rho}\sin\theta_1\cos\varphi_1, ~
\lambda_y=-\frac{3-\eta}{2\sqrt{3}\rho}\sin\theta_1\sin\varphi_1, \\ \nonumber
\lambda_z &=&-\frac{\eta}{\sqrt{3}\rho}\cos\theta_1, ~
\omega_1=-\gamma H_r.
\ee
So, the evolution operator $R(t)$ in the interaction representation is given (in the $|\phi_i\rangle$ basis) by
\be
R(t)=\left(\begin{array}{cccc}
\cos\lambda\Gamma&0&-\frac{\lambda_y+i\lambda_x}{\lambda}\sin\lambda\Gamma&i\frac{\lambda_z}{\lambda}\sin\lambda\Gamma\\
0&\cos\lambda\Gamma&-i\frac{\lambda_z}{\lambda}\sin\lambda\Gamma&\frac{\lambda_y-i\lambda_x}{\lambda}\sin\lambda\Gamma\\
\frac{\lambda_y-i\lambda_x}{\lambda}\sin\lambda\Gamma&-i\frac{\lambda_z}{\lambda}\sin\lambda\Gamma&\cos\lambda\Gamma&0\\
i\frac{\lambda_z}{\lambda}&-\frac{\lambda_y+i\lambda_x}{\lambda}\sin\lambda\Gamma&0&\cos\lambda\Gamma\\
\end{array}\right),
\ee
where 
\be
\lambda^2=\lambda_x^2+\lambda_y^2+\lambda_z^2,~
\Gamma&=&-\gamma H_rt_w.
\ee
We transform $R^i(t)$  to the Schrodinger represenation and in the basis $|\xi_i\rangle$ given by (\ref{basis2}):
(\ref{basis2}),
\be
R(t)=U_0VR^i(t)V^{-1},
\ee
where $V$ is the unitary transformation from basis $|\phi_i\rangle$ to basis $|\xi_i\rangle$.

We can now evaluate, using (\ref{I}), the time-dependence of the signal picked up by the rf coil oriented at random. The result is proportional to 
\be
\label{I-gen}
\langle I(t) \rangle &=&\frac{6\sqrt{3}NP}{(2I+1)k\Theta K^{3/2}}\sin2\sqrt{K}\omega_1t_w\sin^2\sqrt{K}\omega_1t_w\sin\omega(t-2\tau)\{K_1^2\cos(B-D)\omega_0(t-2\tau)\nonumber\\
&&+K_2^2\cos(B+D)\omega_0(t-2\tau)+2K_1K_2[\cos B\omega_0(t-2\tau)\cos D\omega_0t\nonumber\\
&&+\cos B\omega_0t\cos D\omega_0(t-2\tau)-\cos B\omega_0t\cos D\omega_0t]\}.
\ee
Here $K=K_1+K_2 = \lambda^2$. (This expression does not include any of the irreversible decay processes which will be taken into account phenomenologically.)
 Eq. (\ref{I-gen}) gives the results in terms of the four modulation frequencies $B\omega_0, D\omega_0$ and $(B \pm D)\omega_0$, and two coefficients $K_1,K_2$. This is consistent with the energy splittings from which a maximum of 5 different excitation energies are possible, one of which is the fundamental and does not appear because it is rectified by the applied rf frequency. The expressions for $K_1,K_2$ in this general case are very complicated and given in the Appendix.  The amplitude of oscillations are dependent on angles of the external coil $(\theta_0, \phi_0)$ and the frequencies depend on the angles $(\theta_1,\ phi_1)$ of the internal fields with respect to the crystalline axis. These are needed only for interpreting experiments in single-crystals  with arbitrary orientation with respect to the rf field. 

If we consider c-axis oriented samples with rf coil in the x-y plane, there are some simplifications in the coefficients in (\ref{I-gen}). The result for the echo envelope amplitude, i.e. $<I(2\tau)>$ in this case is 
\be
\label{echoamp}
<I(2\tau)>=\frac{\sin(2\lambda\omega_1t_w)\sin^2(\lambda\omega_1t_w)}{\lambda^3}\{K_1^2+K_2^2+ \\ \nonumber
2K_1K_2[\cos(2D\omega_0\tau)+\cos(2B\omega_0\tau)-\cos(2D\omega_0\tau)\cos(2B\omega_0\tau)]\}.
\ee
This result still depends on the angles $\theta_0$, and $(\varphi_0 - \varphi_1)$ through the dependence of $K_{1,2}$ on these angles:
 \be
\label{K1K2}
K_1^2+K_2^2=\lambda^4\{\frac{(\rho^2-\eta^2-1+2\eta\cos\varphi_0)^2+(\rho^2+\eta^2-1)^2\sin^22\varphi_0}{16\rho^2BD(2\rho B+(2+\rho)\cos\theta_0)(2\rho D+(2-\rho)\cos\theta_0))}\sin^4\theta_0\nonumber\\
+\frac{(2+\rho)^2D^2+(2-\rho)^2B^2}{16\rho^2B^2D^2}\cos^2\theta_0+\frac{(2+\rho)^2D^2+(2-\rho)^2B^2}{4\rho^2(2+\rho)^2D^2)}\},\nonumber\\
2K_1K_2=\lambda^4\{-\frac{(\rho^2-\eta^2-1+2\eta\cos\varphi_0)^2+(\rho^2+\eta^2-1)^2\sin^22\varphi_0}{16\rho^2BD(2\rho B+(2+\rho)\cos\theta_0)(2\rho D+(2-\rho)\cos\theta_0))}\sin^4\theta_0\nonumber\\
+\frac{16\rho^2B^2D^2-[(2+\rho)^2D^2+(2-\rho)^2B^2]\cos^2\theta_0}{16\rho^2B^2D^2}-\frac{(2+\rho)^2D^2+(2-\rho)^2B^2}{4\rho^2(2+\rho)^2D^2)}\}.
\ee
\noindent
For orthorhombic samples aligned along the c-axis but powdered in the a-b plane, and r-f field in the basal plane, one must average  Eq. (\ref{echoamp}) over $(\phi_1 -\phi_0)$. This is done numerically in the results which we will display below.
 
\section{Results for Tetragonal crystals with Zeeman field  in the c-axis and in the basal planes}

We now check that our general results reduce to the classic results \cite{Das1, bloom} for tetragonal symmetry, i.e. $\eta =0, \rho =1$, and for the case that the Zeeman field is applied along the c-axis, i.e. $\theta_0=0$ or in the basal plane, i.e. $\theta_0 =\pi/2$.\\

{\it Tetragonal with Zeeman field along c-axis}: In this case, $B=1/2, D=3/2$. Inserting in (\ref{K1K2}) gives $K_1K_2 = 0$.  It follows  from Eq. (\ref{echoamp}) that in this case, the amplitude of the periodic term is 0, in agreement with \onlinecite{Das1} but  contrary to the expressions in References (\onlinecite{Lombardi1, Mali}). The null result is easy to understand. In this case the eigen-functions with the uniform field continue to be eigenstates of $I_z$. So independent transitions are induced between states differing in $I_z$ by 1 if the rf field is in a direction transverse to the z (i.e. c)-axis. There is no mixing of the transitions and therefore no modulation.\\

{\it Tetragonal with Zeeman field anywhere in the basal plane}: In this case, we get from (\ref{BD}) that $B=0, D =2$. In this case the amplitude factor $K_1K_2 \ne 0$. But  a look at Eq.(\ref{echoamp}) shows that the the periodic terms in (\ref{echoamp}) cancel each other. This is also an ancient result \cite{bloom}, but again again contrary to Ref. ((\onlinecite{Lombardi1, Mali}). The explanation of the correct result is that with $B =0$, the states of $I_z = \pm 1/2$ remain degenerate and therefore any linear combination of them with arbitrary phase factors is allowed. So a proper calculation of transitions to the the higher states which are linear combinations of $I_z = \pm 3/2$, the 
mixing terms must cancel. \\

\subsection{Comparison with Experiments on Tetragonal Crystals}
The old results, derived anew here, are exact for atoms sitting in sites of tetragonal symmetry and for rf field polarized in the basal planes. So they should bear comparison with well defined experiments meeting the specified conditions.
Spin-echo modulation experiments \cite{Lombardi2} have been done on the Ba nuclei the insulating AFM, $YBa_2Cu_3O_{6.05}$. This is a tetragonal crystal and Ba sit on sites of tetragonal symmetry. The direction of the internal magnetic field due to the AFM ordering of the spins on $Cu^{2+}$ at the Ba sites is in the basal plane. The experiment was done on an un-oriented powdered sample. But since this is an internal field, this should not matter for any given crystallite. No modulation should therefore be observed. However, modulations were indeed observed and ascribed to a single frequency generated by the internal field. This field was about 1/2 of what is expected from the known magnetic moment on Cu in this compound. The finite signal observed could only be from Ba sitting at sites of lower symmetry than in the interior of the crystallite, for example from those at the boundaries of crystallites. But this can only be a very tiny effect in amplitude. Moreover,
how a single frequency could be observed is a mystery.

\begin{figure}
\includegraphics[width=0.5\columnwidth]{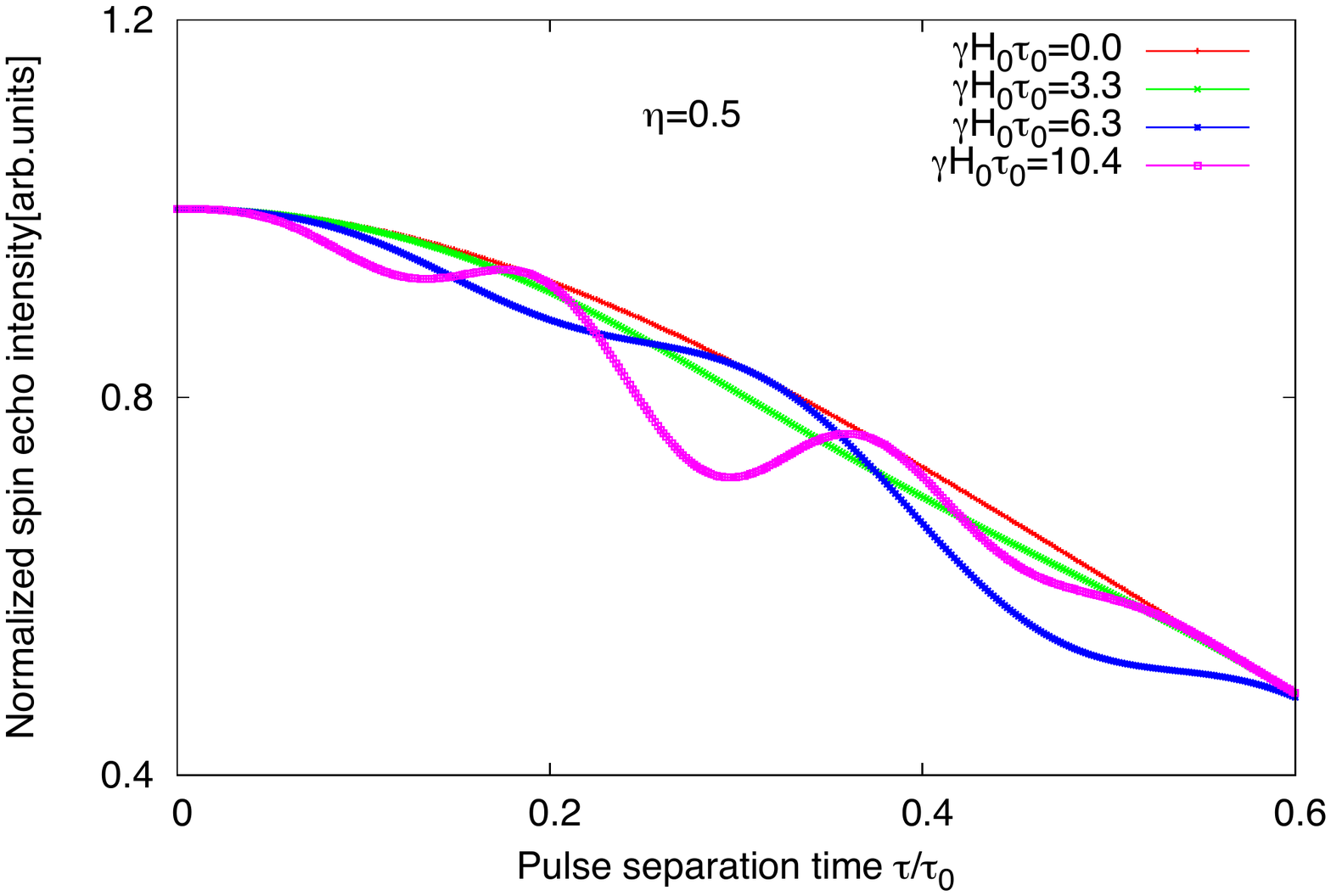} 
\caption{The quadrupolar echo plotted as a function of  the time between pulses normalized to a phenomenological decay rate $\tau_0$ for a fixed asymmetry parameter $\eta$ and various values os an external Zeeman field applied along the c-axis with frequency also normalized with respect to the decay rate.}
\label{Fig:external(parral,B2.3G,varing eta)}
\end{figure}

\begin{figure}
\includegraphics[width=0.5\columnwidth]{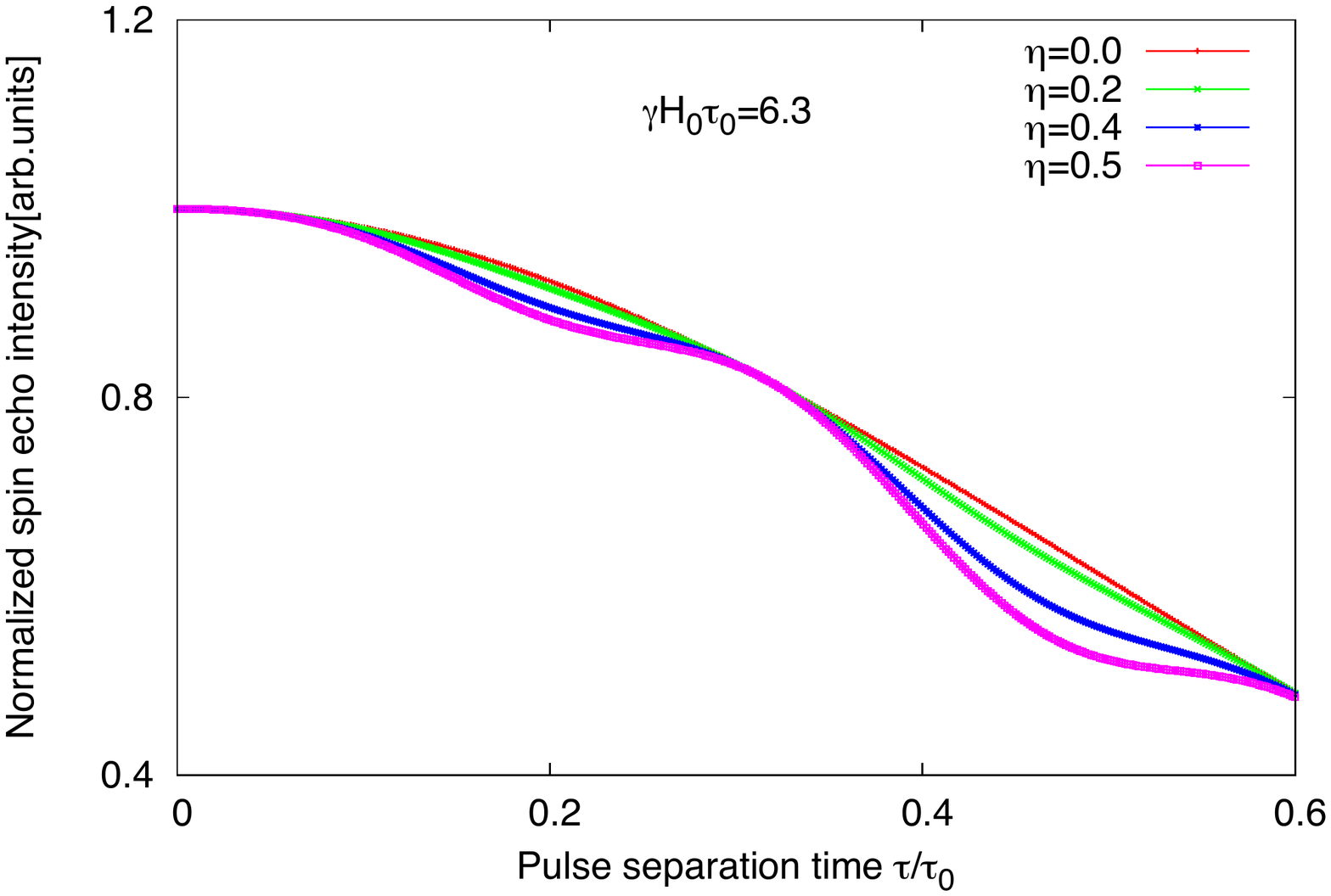} 
\caption{The quadrupolar echo plotted as a function of  the time between pulses normalized to a phenomenological decay rate $\tau_0$ for a fixed external Zeeman field in the basal plane with frequency also normalized with respect to the decay rate for various values of the orthorhombic parameter $\eta$.}
\label{Fig:external(perp,B=2.3G,varing eta)}
\end{figure}

\begin{figure}
\includegraphics[width=0.5\columnwidth]{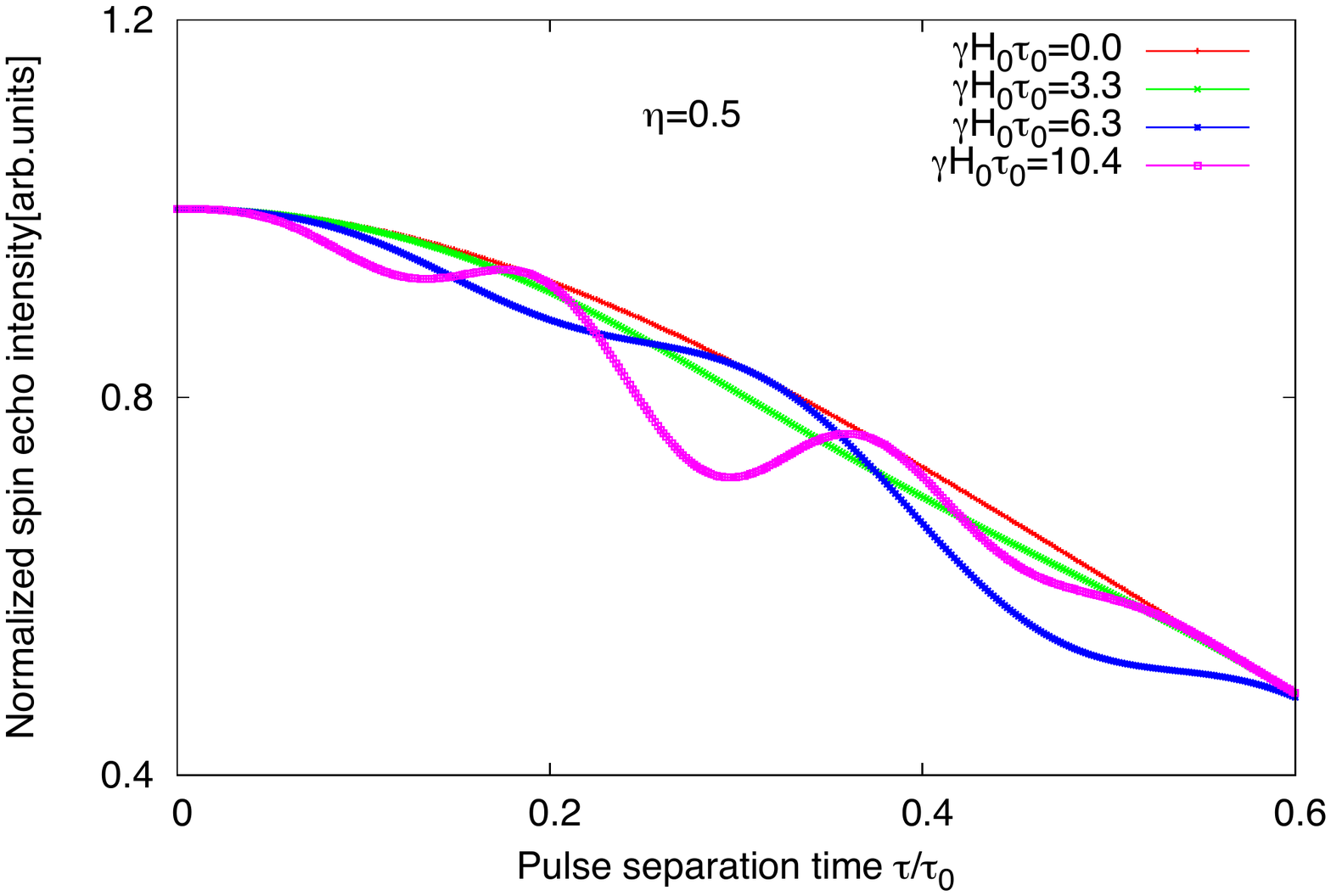} 
\caption{The quadrupolar echo plotted as a function of  the time between pulses normalized to a phenomenological decay rate $\tau_0$ for a fixed external Zeeman field in the basal plane with frequency also normalized with respect to the decay rate for various values of the orthorhombic parameter $\eta$.}
\label{Fig:external(parral,eta=0.5,varing B)}
\end{figure}

\begin{figure}
\includegraphics[width=0.5\columnwidth]{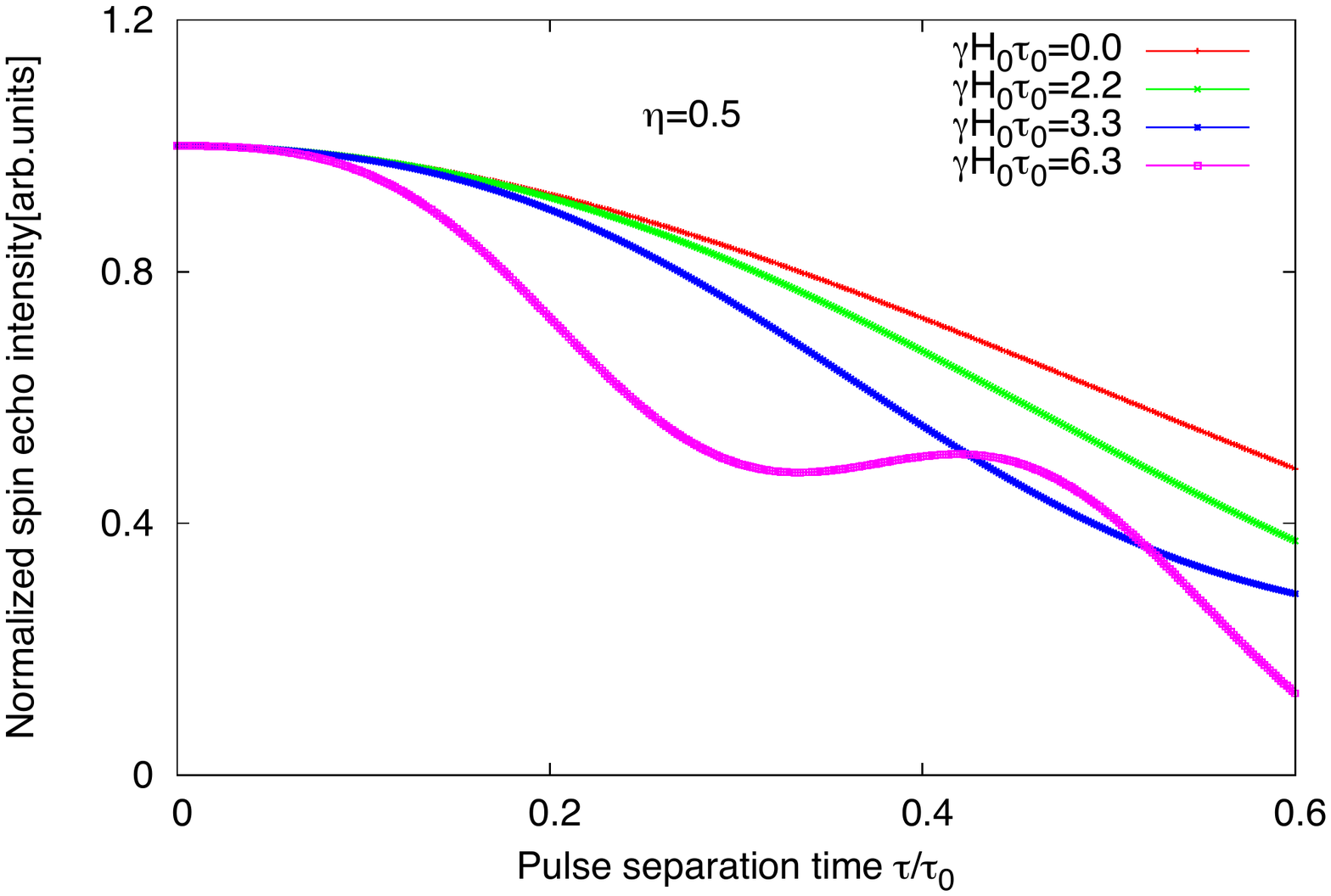} 
\caption{The quadrupolar echo plotted as a function of  the time between pulses normalized to a phenomenological decay rate $\tau_0$ for a fixed asymmetry parameter $\eta$ and varying external Zeeman field in the basal plane with frequency also normalized with respect to the decay rate.}
\label{Fig:external(perp,eta=0.5,varing B)}
\end{figure}

\section{Modulation Echoes for  Orthorhombic crystals}
\subsection{Externally Applied Zeeman fields} 
In this case, the results are complicated enough that numerical evaluation of Eq.(\ref{echoamp}) is required.  The results for rf- field in the basal plane are averaged over the angle  $\phi_1$, as appropriate for a powdered sample aligned along the c-axis. The time-dependence of the modulation is shown in units of a phenomenological decay rate $\tau_0$ for various values of the orthorhombic parameter $\eta$ for a fixed Zeeman field along the c-axis in Fig (\ref{Fig:external(parral,B2.3G,varing eta)})
 and in the basal plane (\ref{Fig:external(perp,B=2.3G,varing eta)}), and for a fixed $\eta$ for various values of the internal Zeeman fields with field along the c-axis, Fig.(\ref{Fig:external(parral,eta=0.5,varing B)}), and in the basal plane, Fig. (\ref{Fig:external(perp,eta=0.5,varing B)}). As expected, the results strongly depend on $\eta$. Typically, modulations are only observable when $\omega_0\tau_0$ is comparable to $\eta$. In the experiments done in orthorhombic $YBa_2Cu_4O_8$, the characteristic decay time $\tau_0 \approx 0.9 msec.$. To help read the figures, it is useful to know that in this case $\gamma H_0 \tau_0 =1$, corresponds to a field $H_0 \approx 0.04 mTesla$. 

\subsection{Comparison of Calculations with Experiments with fixed external fields}
Experiments are done on Ba nuclei in metallic under-doped cuprate $YBa_2Cu_4O_8$ on a c-axis oriented poly-crystalline sample. The single-crystal is orthorhombic and Ba sit at a site of orthorhombic symmetry. A fixed external magnetic field was applied either along the c-axis or the basal plane. Modulations were observed which we ought to be able to compare with results in Figs. (2)-(5). We note that for the orthorhombic parameter $\eta \lesssim 0.2$, no oscillations are observable if $\gamma H_0 \tau_0 \lesssim 10$, which corresponds to a field of about 0.4 mTesla. However for larger $\eta$ oscillations are indeed observable. A surprisingly large value of $\eta \approx 0.56$ has been deduced in experiments \cite{Lombardi2 }. If we accept such a value, oscillations should indeed be observed for fields either in the basal plane or along the c-axis with value $\gamma H_0\tau_0$ of $O(10)$. Such are indeed the magnitudes of the fields applied. But the results are fitted to a single frequency, whereas our results show that four different frequencies with similar amplitudes ought to be observed. We have no explanation of this discrepancy of the experiments with our results which for the specified conditions are exact. 

\begin{figure}
\includegraphics[width=0.5\columnwidth]{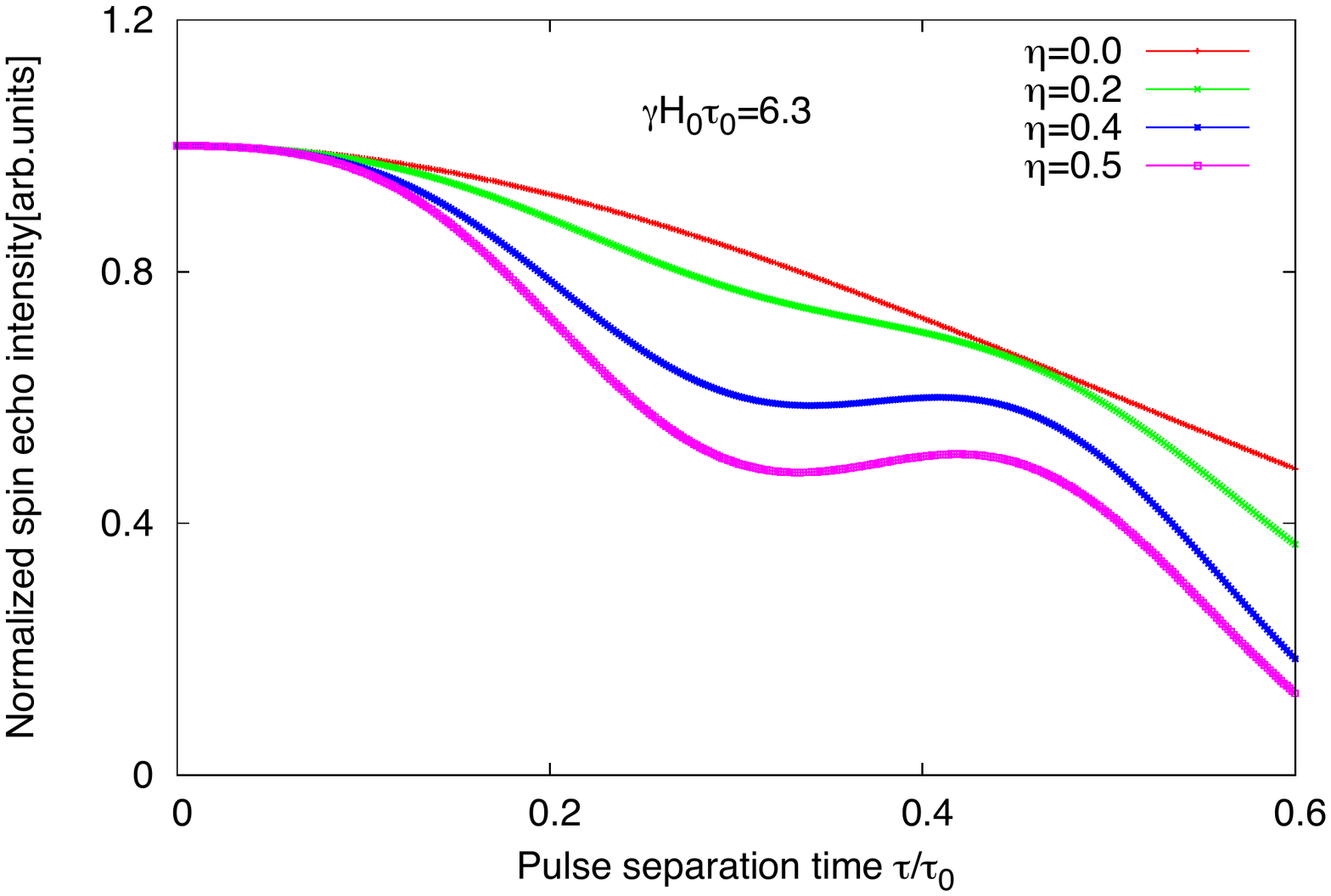} 
\caption{The quadrupolar echo plotted as a function of  the time between pulses normalized to a phenomenological decay rate $\tau_0$ for the internal basal Zeeman field intrinsic to the sample specified in the figure for  various values of the orthorhombic parameter $\eta$.}
\label{Fig:intrinsic(B=2.3G,varing eta)} 
\end{figure}

\begin{figure}
\includegraphics[width=0.5\columnwidth]{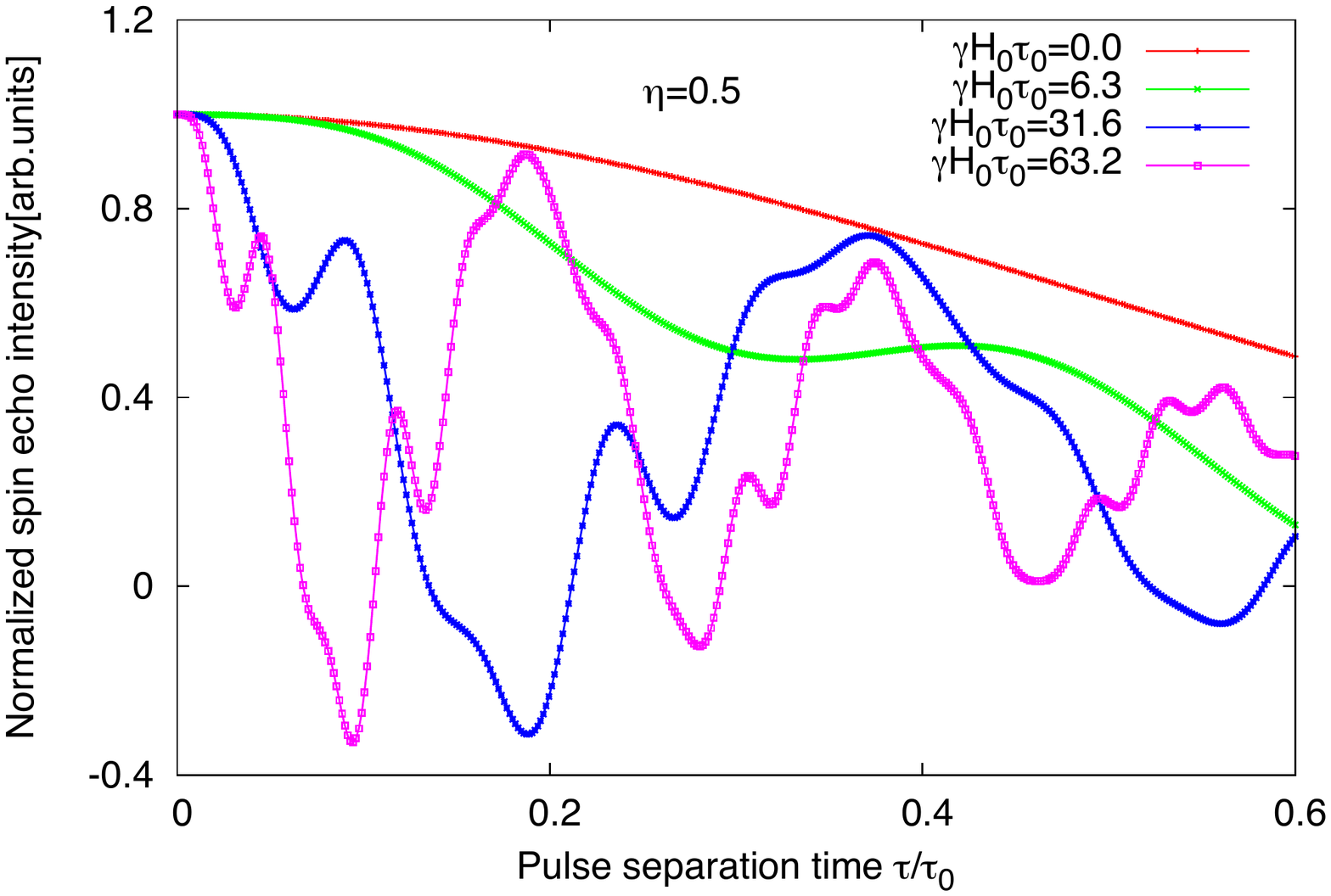} 
\caption{The quadrupolar echo plotted as a function of  the time between pulses normalized to a phenomenological decay rate $\tau_0$ for  several  intrinsic Zeeman fields in the basal plane with frequency normalized with respect to the decay rate for a large fixed value of the orthorhombic parameter $\eta =0.5$.}
\label{Fig:intrinsic(eta=0.5,varing B)}
\end{figure}

\subsection{Modulation of Echoes expected in static Loop Ordered Phase in Cuprates}
If the loop order were static on the scale of $O(10^{-5})$ secs, internal fields should in principle be observable at Ba nuclei, which do not sit on a site of high enough symmetry to cancel the  the fringe fields of the loop currents. Different domains of loop order are expected to have fields in the $\pm(x\pm y))$ directions. These have to be averaged in a given crystallite. Then there should be an average over the angle $\phi_0$, i.e of the coil with respect to the random orientation of the crystallite. We have done such calculations and find that 
 for $\eta = 0.56$, oscillations should be seen if the loop order is static. Results for various values of $\eta$ for a fixed magnitude of internal field normalized to the relaxation rate are presented in Fig.(\ref{Fig:intrinsic(B=2.3G,varing eta)}). Results for fixed $\eta$ and various internal fields are presented in Fig.(\ref{Fig:intrinsic(eta=0.5,varing B)}). Fields of $O(50)$-Gauss are expected. For $\eta \approx 0.5$, they should have been observed if the order was truly static. As mentioned, a way of resolving the discrepancy between the neutron diffraction experiments which have a time scale of $O(10^{-11})$ secs and NQR experiments which have a scale of $O(10^{-5})$ secs. is that the order has fluctuations at a scale intermediate between two such wide varying scales. However, it should also be borne in mind that  the experiments which we have tried to understand do not give the  answers in agreement with the classic theoretical results even for external known fields. 
 
\noindent
{\it Acknowledgements}: CMV's work is supported by NSF grant DMR 1206298.
\section{Appendices}
\noindent
{\bf A-1 Representation of the Quadrupole Operator in an Orthorhombic Symmetry}: \\

The suitable basis for a quadrupole ${\bf I}$ in an orthorhombic crystal field is given by Eq. (\ref{basis1}). Given this basis, it is straightforward to calculate that
\be
I_x&=&\left(\begin{array}{cccc}
0&\frac{\rho+\eta-1}{2\rho}&\frac{3+\eta}{2\sqrt{3}\rho}&0\\
\frac{\rho+\eta-1}{2\rho}&0&0&\frac{3+\eta}{2\sqrt{3}\rho}\\
\frac{3+\eta}{2\sqrt{3}\rho}&0&0&\frac{\rho-\eta+1}{2\rho}\\
0&\frac{3+\eta}{2\sqrt{3}\rho}&\frac{\rho-\eta+1}{2\rho}&0
\end{array}\right),\nonumber\\
I_y&=&i\left(\begin{array}{cccc}
0&\frac{\rho-\eta-1}{2\rho}&\frac{3-\eta}{2\sqrt{3}\rho}&0\\
-\frac{\rho-\eta-1}{2\rho}&0&0&\frac{\eta-3}{2\sqrt{3}\rho}\\
\frac{\eta-3}{2\sqrt{3}\rho}&0&0&\frac{\rho+\eta+1}{2\rho}\\
0&\frac{3-\eta}{2\sqrt{3}\rho}&-\frac{\rho+\eta+1}{2\rho}&0
\end{array}\right),\nonumber\\
I_z&=&\left(\begin{array}{cccc}
-\frac{\rho+2}{2\rho}&0&0&\frac{\eta}{\sqrt{3}\rho}\\
0&\frac{\rho+2}{2\rho}&-\frac{\eta}{\sqrt{3}\rho}&0\\
0&-\frac{\eta}{\sqrt{3}\rho}&\frac{\rho-2}{2\rho}&0\\
\frac{\eta}{\sqrt{3}\rho}&0&0&\frac{2-\rho}{2\rho}
\end{array}\right)
\ee

\noindent
{\bf A-2 Perturbative Eigenvalues, Eigenfunctions and representation of Quadrupole Operator with applied Zeeman Field}:\\

Consider the full Hamiltonian given by (\ref{fullH})  which include the Zeeman perturbation. For $H_Q\gg H_Z$, the wavefunctions and the energy levels can be obtained by treating the Zeeman term as a perturbation. The energy levels are given by
\be
E_{2,1}&=&-E_Q\pm\frac{\hbar\gamma H_0}{2\rho}[(2-\rho)^2\cos^2\theta_0+(\rho+1-\eta)^2\sin^2\theta_0\cos^2\varphi_0+(\rho+1+\eta)^2\sin^2\theta_0\sin^2\varphi_0]^{1/2},\nonumber\\
E_{4,3}&=&E_Q\pm\frac{\hbar\gamma H_0}{2\rho}[(2+\rho)^2\cos^2\theta_0+(\rho-1+\eta)^2\sin^2\theta_0\cos^2\varphi_0+(\rho-1-\eta)^2\sin^2\theta_0\sin^2\varphi_0]^{1/2}
\ee
where $E_Q=\frac{e^2qQ}{4}\rho$, $\theta_0$ and $\varphi_0$ are the direction of the Zeeman field. Set 
\be
D&=&\frac{[(2-\rho)^2\cos^2\theta_0+(\rho+1-\eta)^2\sin^2\theta_0\cos^2\varphi_0+(\rho+1+\eta)^2\sin^2\theta_0\sin^2\varphi_0]^{1/2}}{2\rho},\nonumber\\
B&=&\frac{[(2+\rho)^2\cos^2\theta_0+(\rho-1+\eta)^2\sin^2\theta_0\cos^2\varphi_0+(\rho-1-\eta)^2\sin^2\theta_0\sin^2\varphi_0]^{1/2}}{2\rho}
\ee
we get 
\be
E_{2,1}&=&-E_Q\pm\hbar D\omega_0,\nonumber\\
E_{4,3}&=&E_Q\pm\hbar B\omega_0
\ee
where $\omega_0=\gamma H_0$. The first order orthonormal eigenfunctions are of the form,
\be
|\xi_1\rangle&=&-d^\ast|\phi_2\rangle+c^\ast|\phi_1\rangle,\nonumber\\
|\xi_2\rangle&=&c|\phi_2\rangle+d|\phi_1\rangle,\nonumber\\
|\xi_3\rangle&=&-b^\ast|\phi_4\rangle+a^\ast|\phi_3\rangle,\nonumber\\
|\xi_4\rangle&=&a|\phi_4\rangle+b|\phi_3\rangle
\ee
where the coefficients $a$ and $b$ satisfy the nonlinear equation,
\be
a^\ast b^\ast+ab&=&-\frac{(\rho-1+\eta)\sin\theta_0\cos\varphi_0}{2\rho B}=-\alpha_1,\nonumber\\
ab-a^\ast b^\ast&=&-i\frac{(\rho-1-\eta)\sin\theta_0\sin\varphi_0}{2\rho B}=-i\beta_1,\nonumber\\
aa^\ast-bb^\ast&=&-\frac{(2+\rho)\cos\theta_0}{2\rho B}=-\gamma_1,\nonumber\\
aa^\ast+bb^\ast&=&1.
\ee
 $c$ and $d$ satisfy
\be
c^\ast d^\ast+c^\ast d^\ast&=&-\frac{(\rho+1-\eta)\sin\theta_0\cos\varphi_0}{2\rho D}=-\alpha_2,\nonumber\\
cd-c^\ast d^\ast&=&-i\frac{(\rho+1+\eta)\sin\theta_0\sin\varphi_0}{2\rho D}=-i\beta_2,\nonumber\\
cc^\ast-dd^\ast&=&-\frac{(2-\rho)\cos\theta_0}{2\rho D}=-\gamma_2,\nonumber\\
cc^\ast+dd^\ast&=&1
\ee
The above equations can not fix the coefficients; we choose $a, c$ to be real, i.e. choose a particular gauge. This problem  is encountered very often when one diagonalizes a matrix. After fixing this arbitrariness, we get
\be
a=\sqrt{\frac{1-\gamma_1}{2}},~
b=-\frac{\alpha_1+i\beta_1}{\sqrt{2(1-\gamma_1)}},~
c=\sqrt{\frac{1-\gamma_2}{2}},~
d=-\frac{\alpha_2+i\beta_2}{\sqrt{2(1-\gamma_2)}}.
\ee
In this basis, the spin operator $I_x,I_y,I_z$ can be written as(the order is $|\xi_4\rangle,|\xi_3\rangle,|\xi_2\rangle,|\xi_1\rangle$)
\be
I_x&=&\left(\begin{array}{cccc}
\frac{\rho+\eta-1}{2\rho}(a^\ast b+b^\ast a)&\frac{\rho+\eta-1}{2\rho}((a^\ast)^2 -(b^\ast)^2)&\frac{3+\eta}{2\sqrt{3}\rho}(a^\ast c+b^\ast d)&\frac{3+\eta}{2\sqrt{3}\rho}(c^\ast b^\ast-a^\ast d^\ast)\\
\frac{\rho+\eta-1}{2\rho}((a^\ast)^2 -(b^\ast)^2)&-\frac{\rho+\eta-1}{2\rho}(a^\ast b+b^\ast a)&\frac{3+\eta}{2\sqrt{3}\rho}(ad-bc)&\frac{3+\eta}{2\sqrt{3}\rho}(ac^\ast+bd^\ast)\\
\frac{3+\eta}{2\sqrt{3}\rho}(ac^\ast+bd^\ast)&\frac{3+\eta}{2\sqrt{3}\rho}(a^\ast d^\ast-b^\ast c^\ast)&\frac{\rho-\eta+1}{2\rho}(c^\ast d+d^\ast c)&\frac{\rho-\eta+1}{2\rho}((c^\ast)^2-(d^\ast)^2)\\
\frac{3+\eta}{2\sqrt{3}\rho}(cb-ad)&\frac{3+\eta}{2\sqrt{3}\rho}(ca^\ast+db^\ast)&\frac{\rho-\eta+1}{2\rho}((c^\ast)^2-(d^\ast)^2)&-\frac{\rho-\eta+1}{2\rho}(c^\ast d+d^\ast c)
\end{array}\right),\nonumber\\
I_y&=&i\left(\begin{array}{cccc}
\frac{\rho-\eta-1}{2\rho}(ba^\ast-ab^\ast)&\frac{\rho-\eta-1}{2\rho}((a^\ast)^2+(b^\ast)^2)&\frac{3-\eta}{2\sqrt{3}\rho}(ca^\ast-db^\ast)&-\frac{3-\eta}{2\sqrt{3}\rho}(a^\ast d^\ast+b^\ast c^\ast)\\
-\frac{\rho-\eta-1}{2\rho}((a^\ast)^2+(b^\ast)^2)&\frac{\rho-\eta-1}{2\rho}(ab^\ast-ba^\ast)&-\frac{3-\eta}{2\sqrt{3}\rho}(bc+ad)&\frac{3-\eta}{2\sqrt{3}\rho}(bd^\ast-ac^\ast)\\
\frac{3-\eta}{2\sqrt{3}\rho}(bd^\ast-ac^\ast)&\frac{3-\eta}{2\sqrt{3}\rho}(db^\ast-ca^\ast)&\frac{\rho+\eta+1}{2\rho}(dc^\ast-cd^\ast)&\frac{\rho+\eta+1}{2\rho}((c^\ast)^2+(d^\ast)^2)\\
\frac{3-\eta}{2\sqrt{3}\rho}(ad+bc)&\frac{3-\eta}{2\sqrt{3}\rho}(ca^\ast-db^\ast)&-\frac{\rho+\eta+1}{2\rho}((c^\ast)^2+(d^\ast)^2)&\frac{\rho+\eta+1}{2\rho}(cd^\ast-dc^\ast)
\end{array}\right),\nonumber\\
I_z&=&\left(\begin{array}{cccc}
\frac{\rho+2}{2\rho}(|b|^2-|a|^2)&\frac{\rho+2}{2\rho}2a^\ast b^\ast&\frac{\eta}{\sqrt{3}\rho}(da^\ast-cb^\ast)&\frac{\eta}{\sqrt{3}\rho}(a^\ast c^\ast+b^\ast d^\ast)\\
\frac{\rho+2}{2\rho}2ab&\frac{\rho+2}{2\rho}(|a|^2-|b|^2)&-\frac{\eta}{\sqrt{3}\rho}(ac+bd)&\frac{\eta}{\sqrt{3}\rho}(ad^\ast-bc^\ast)\\
\frac{\eta}{\sqrt{3}\rho}(ad^\ast-bc^\ast)&-\frac{\eta}{\sqrt{3}\rho}(b^\ast d^\ast+a^\ast c^\ast)&\frac{\rho-2}{2\rho}(|c|^2-|d|^2)&-\frac{\rho-2}{2\rho}2c^\ast d^\ast\\
\frac{\eta}{\sqrt{3}\rho}(ac+bd)&\frac{\eta}{\sqrt{3}\rho}(da^\ast-cb^\ast)&-\frac{\rho-2}{2\rho}2cd&\frac{\rho-2}{2\rho}(|d|^2-|c|^2)
\end{array}\right)
\ee

{\bf A-3 Spin-Echo Modulations}
Using the representation of the quadrupole operator ${\bf I}$ in the last section and following Eq. (\ref{I}), it is lengthy but straightforward to calculate that the spin echo envelope modulations,
\be
\langle I\rangle&=&\frac{6\sqrt{3}NP}{(2I+1)k\Theta K^{3/2}}\sin2\sqrt{K}\omega_1t_w\sin^2\sqrt{K}\omega_1t_w\sin\omega(t-2\tau)\{K_1^2\cos(B-D)\omega_0(t-2\tau)\nonumber\\
&&+K_2^2\cos(B+D)\omega_0(t-2\tau)+2K_1K_2[\cos B\omega_0(t-2\tau)\cos D\omega_0t\nonumber\\
&&+\cos B\omega_0t\cos D\omega_0(t-2\tau)-\cos B\omega_0t\cos D\omega_0t]\}
\ee
where
\be
K&=&K_1+K_2=[\frac{\eta^2}{3}+\frac{1}{4}(3-\eta^2+2\eta\cos2\varphi_1)\sin^2\theta_1]/\rho^2=\lambda^2,\nonumber\\
2K_1K_2&=&\lambda^3(A_1P_1+B_1Q_1+C_1R_1+D_1S_1),\nonumber\\
K_1^2+K_2^2&=&\lambda^3(E_1P_1+F_1Q_1+G_1R_1+H_1S_1),\nonumber\\
\rho&=&[1+\eta^2/3]^{1/2},\nonumber\\
B&=&[(2+\rho)^2\cos^2\theta_0+(\rho-1+\eta)^2\sin^2\theta_0\cos^2\varphi_0+(\rho-1-\eta)^2\sin^2\theta_0\sin^2\varphi_0]^{1/2}/2\rho,\nonumber\\
D&=&[(2-\rho)^2\cos^2\theta_0+(\rho+1-\eta)^2\sin^2\theta_0\cos^2\varphi_0+(\rho+1+\eta)^2\sin^2\theta_0\sin^2\varphi_0]^{1/2}/2\rho
\ee
where
\be
P_1&=&(I_x)_{24}\sin\theta_1\cos\varphi_1+(I_y)_{24}\sin\theta_1\sin\varphi_1+(I_z)_{24}\cos\theta_1,\nonumber\\
Q_1&=&(I_x)_{23}\sin\theta_1\cos\varphi_1+(I_y)_{23}\sin\theta_1\sin\varphi_1+(I_z)_{23}\cos\theta_1,\nonumber\\
R_1&=&(I_x)_{14}\sin\theta_1\cos\varphi_1+(I_y)_{14}\sin\theta_1\sin\varphi_1+(I_z)_{14}\cos\theta_1,\nonumber\\
S_1&=&(I_x)_{13}\sin\theta_1\cos\varphi_1+(I_y)_{13}\sin\theta_1\sin\varphi_1+(I_z)_{13}\cos\theta_1
\ee
and 
\be
A_1&=&+iZ_1Z_2Z_2^\ast, E_1=+iZ_1Z_1Z_1^\ast,\nonumber\\
B_1&=&-iZ_1Z_1^\ast Z_2, F_1=-iZ_2Z_2Z_2^\ast,\nonumber\\
C_1&=&-iZ_1Z_1Z_2^\ast, G_1=-iZ_2Z_2^\ast Z_2^\ast,\nonumber\\
D_1&=&-iZ_1^\ast Z_2Z_2, H_1=-iZ_1Z_1^\ast Z_1^\ast
\ee
Here
\be
Z_1&=&(-ac^\ast f-ibc^\ast g+iad^\ast g+bd^\ast f^\ast),\nonumber\\
Z_2&=&(adf+ibdg+iacg+bcf^\ast),\nonumber\\
f&=&\frac{\lambda_y+i\lambda_x}{\lambda};g=\frac{\lambda_z}{\lambda}
\ee
After some algebra, we get
\be
E_1P_1+F_1Q_1+G_1R_1+H_1S_1&=&\lambda Z_1Z_1^\ast\{Z_1(-ac^\ast f^\ast+bd^\ast f-iad^\ast g+ibc^\ast g)
+Z_1^\ast(-a^\ast cf+b^\ast df^\ast+ia^\ast dg-ib^\ast cg)\}\nonumber\\
&-&\lambda Z_2Z_2^\ast\{Z_2(-adf^\ast-bcf+iacg+ibdg)+Z_2^\ast(-a^\ast d^\ast f-b^\ast c^\ast f^\ast-ia^\ast c^\ast g-ib^\ast d^\ast g)\},\nonumber\\
A_1P_1+B_1Q_1+C_1R_1+D_1S_1&=&\lambda Z_2\{Z_1Z_2^\ast(-ac^\ast f^\ast+bd^\ast f-iad^\ast g+ibc^\ast g)+Z_1^\ast Z_2(-a^\ast cf+b^\ast df^\ast+ia^\ast dg-ib^\ast cg)\}\nonumber\\
&+&\lambda Z_1\{Z_2Z_1^\ast(-adf^\ast-bcf+iacg+ibdg)+Z_1Z_2^\ast(-a^\ast d^\ast f-b^\ast c^\ast f^\ast-ia^\ast c^\ast g-ib^\ast d^\ast g)\}\nonumber\\
\ee
Set 
\be
Z_3&=&-ac^\ast f^\ast+bd^\ast f-iad^\ast g+ibc^\ast g,\nonumber\\
Z_4&=&-adf^\ast-bcf+iacg+ibdg
\ee
so we have
\be
E_1P_1+F_1Q_1+G_1R_1+H_1S_1&=&\lambda Z_1Z_1^\ast\{Z_1Z_3+Z_1^\ast Z_3^\ast\}-\lambda Z_2Z_2^\ast\{Z_2Z_4+Z_2^\ast Z_4^\ast\},\nonumber\\
A_1P_1+B_1Q_1+C_1R_1+D_1S_1&=&\lambda Z_2\{Z_1Z_2^\ast Z_3+Z_1^\ast Z_2Z_3^\ast\}+\lambda Z_1\{Z_2Z_1^\ast Z_4+Z_1Z_2^\ast Z_4^\ast\}
\ee
The above expressions are very complicated but they can be simplified if, as is common in the experiments, the coil field lies in the $x-y$ plane. In that case, we set $\theta_1=\frac{\pi}{2}$, so that $\lambda_z=0$ or $g=0$
Now, we have
\be
Z_1&=&-ac^\ast f+bd^\ast f^\ast,\nonumber\\
Z_2&=&adf+bcf^\ast,\nonumber\\
Z_3&=&-ac^\ast f^\ast+bd^\ast f,\nonumber\\
Z_4&=&-adf^\ast-bcf
\ee
Plugging in $a,b,c$, and $d$,we have
\be
Z_1Z_1^\ast&=&\frac{1}{2\lambda^2}\{(\gamma_1+\gamma_2)\lambda^2-(\alpha_1\alpha_2+\beta_1\beta_2)(\lambda_y^2-\lambda_x^2)+2(\alpha_1\beta_2-\beta_1\alpha_2)\lambda_x\lambda_y\},\nonumber\\
Z_2Z_2^\ast&=&\frac{1}{2\lambda^2}\{(1-\gamma_1\gamma_2)\lambda^2+(\alpha_1\alpha_2+\beta_1\beta_2)(\lambda_y^2-\lambda_x^2)-2(\alpha_1\beta_2-\beta_1\alpha_2)\lambda_x\lambda_y\},\nonumber\\
Z_1Z_3+Z_1^\ast Z_3^\ast&=&\frac{1}{2\lambda^2}\{\frac{(\alpha_1\alpha_2+\beta_1\beta_2)^2-(\alpha_1\beta_2-\beta_1\alpha_2)^2-(1-\gamma_1)^2(1-\gamma_2)^2}{(1-\gamma_1)(1-\gamma_2)}\lambda^2-2(\alpha_1\alpha_2+\beta_1\beta_2)(\lambda_y^2-\lambda_x^2)\},\nonumber\\
Z_2Z_4+Z_2^\ast Z_4^\ast&=&\frac{1}{2\lambda^2}\{\frac{(\alpha_1^2-\beta_1^2)(1-\gamma_2)^2-(\alpha_2^2-\beta_2^2)(1-\gamma_1)^2}{(1-\gamma_1)(1-\gamma_2)}\lambda^2+2(\alpha_1\alpha_2-\beta_1\beta_2)(\lambda_y^2-\lambda_x^2)\}
\ee
After some algebra, we get
\be
K_1^2+K_2^2&=&\lambda^4\{\frac{(\alpha_1\alpha_2+\beta_1\beta_2)^2+(\alpha_1\beta_2+\beta_1\alpha_2)^2}{4(1+\gamma_1)(1+\gamma_2)}+\frac{(\gamma_1^2+\gamma_2^2)(1+\gamma_1^2)}{4\gamma_1^2}\},\nonumber\\
2K_1K_2&=&\lambda^4\{\frac{4\gamma_1^2-(\gamma_1^2+\gamma_2^2)(1+\gamma_1^2)}{4\gamma_1^2}-\frac{(\alpha_1\alpha_2+\beta_1\beta_2)^2+(\alpha_1\beta_2+\beta_1\alpha_2)^2}{4(1+\gamma_1)(1+\gamma_2)}\}
\ee
The expression at $t = 2\tau$, which gives the modulation amplitude is given in Eq. (\ref{echo amp}).

\end{document}